# Epoxy-inspired Nonlinear Interface Integrating Monolayer Transition-Metal Dichalcogenides with Linear Plasmonic Nanosieves

*Xuanmiao Hong[1†], Guangwei Hu[2,3†], Wenchao Zhao[1†], Kai Wang[1]\*, Shang Sun[2], Rui Zhu[2,5], Jing Wu[4], Weiwei Liu[1], Loh Kian Ping[6], Andrew Thye Shen Wee[5,7], Bing Wang[1], Andrea Alù[3]\*, Cheng-Wei Qiu[2]\* and Peixiang Lu[1,8]\**

[1]Wuhan National Laboratory for Optoelectronics and School of Physics, Huazhong University of Science and Technology, Wuhan 430074, China

[2]Department of Electrical and Computer Engineering, National University of Singapore, 4 Engineering Drive 3, Singapore 117583, Singapore.

[3]Advanced Science Research Center, City University of New York, New York 10031, USA.

[4]Institute of Materials Research and Engineering, A*STAR (Agency for Science, Technology and Research), 2 Fusionopolis Way, Innovis, #08-03, 138634, Singapore

[5]Department of Physics, National University of Singapore, 2 Science Drive 3, Singapore, 117551 Singapore

[6]Department of Chemistry National University of Singapore 3 Science Drive 3, Singapore 17543, Singapore

[7]Centre for Advanced 2D Materials, National University of Singapore, Block S14, 6 Science Drive 2, Singapore, 117546 Singapore

[8]Hubei Key Laboratory of Optical information and Pattern Recognition, Wuhan Institute of Technology, Wuhan 430205, China

[†]These authors contributed equally: Xuanmiao Hong, Guangwei Hu and Wenchao Zhao

*Corresponding authors:

Kai Wang, kale_wong@hust.edu.cn

Andrea Alù, aalu@gc.cuny.edu

Cheng-Wei Qiu, chengwei.qiu@nus.edu.sg

Peixiang Lu, lupeixiang@hust.edu.cn

# Abstract

Although TMDC monolayers offer giant optical nonlinearity within few-angstrom thickness, it is still elusive to modulate and engineer the wavefront of nonlinear emissions. The grain size of high-quality monolayers also restricts possibilities of imparting inhomogeneous or gradient profiles of phases and amplitudes by classical light sources. Plasmonic nanosieves can support huge field enhancement and precise nonlinear phase control with hundred-nanometer pixel-level resolution, however they suffer from intrinsically weak nonlinear susceptibility. In nature, epoxy represents a strong glue whose magic adhesion comes from the bonding of two intrinsic loose and nonsticky resins. Inspired by the nature of epoxy, we report a multifunctional and powerful nonlinear interfaces via integrating transition-metal dichalcogenide (TMDC) monolayers with linear plasmonic nanosieves, which cannot be accessed by either constituents. We experimentally demonstrate orbital angular momentum (OAM) generation, beam steering, versatile polarization control and holograms and of SH emission, with the proposed nonlinear interfaces. In addition, an effective second-harmonic (SH) nonlinearity $\chi^{(2)}$ of ~25 nm/V is obtained for such systems. This designer platform synergizes the complementary advantages of TMDC monolayer, plasmonic nanosieves, geometric phases, and field enhancement, paving a distinct avenue toward designer, multi-functional, and ultra-compact nonlinear optical devices.

In the past decade, nonlinear signals (second- and third-harmonic generation, SHG and THG) from nanomaterials and artificially structured materials, namely metamaterials or metasurfaces, have attracted extensive attention for their wavelength tunability, coherence and ultrafast response[1-6], which shows promising applications in nanoprobing[1], imaging[2,3] and crystalline detection[4-6]. In particular, the wavefront engineering and controlling of generated nonlinear signals is the most challenging key to realizing integrated, ultra-thin and efficient nonlinear optical devices, ranging from nonlinear beam steering[7,8], nonlinear metalenses[9-11], nonlinear holography[12,13], optical image encoding[14], and generation of nonlinear optical vortex beam[15-17]. Up to now, there have been several pioneering works on nonlinear phase control based on plasmonic metasurfaces, most of which focus on SHG manipulation[9,13-20]. However, their weak second-order effective susceptibility, $\chi^{(2)}$, due to the inherently weak nonlinear response stemming from structural symmetry breaking at the surface[15,21] of plasmonic materials, poses an intrinsic challenge for wider applications. Although SHG can be boosted by orders of magnitude to the level of ~pm/V under resonance conditions, the samples may be easily damaged in the presence of accumulated heat due to the strong absorption. Alternatively, nonlinear hybrid metasurfaces, such as Au/multiple quantum well (MQW), have been proposed instead of traditional plasmonic metasurfaces[22-24]. Previous works report 2~3 orders of magnitude of enhancement following this route[25-28]. Nevertheless, bulky configurations with demanding fabrication requirements make it difficult to design ultrathin and compact multifunctional nonlinear metasurfaces, letting alone the wavefront control of the SHG signals.

On the other hand, emerging 2D materials, such as transition-metal dichalcogenides (TMDC) monolayers, have been revealed with their excellent electronic and optoelectronic properties[29-32]. Of particular interest is the giant nonlinearity of monolayer TMDCs (for example, the chemical composition is $AB_2$, A=Mo, W, Ta, Nb and B=S, Se, Te). Their broken centrosymmetry enables a second-order susceptibility $\chi^{(2)}$ around 1~100 nm/V[33-36], which is much higher than that of many plasmonic metasurfaces (See section 3 in SI for details) and even comparable to the record values reported in Au/MQW hybrid metasurface (~54 nm/V)[22]. However, the atomic thickness of monolayer TMDCs has inherently weak light-matter interactions, and the limited grain size of high-quality TMDC monolayers (usually about tens of micrometers) almost excludes the opportunity of imparting varying amplitude, polarization, phase, and even angular momentum by classical optical light sources such as spatial light modulators due to the pixel size restriction. Paradigm-shift mechanisms should be explored to boost, manipulate and control the SH photons of atomically thin materials, simultaneously.

The nature provides the hint. It is well known that epoxy is a type of strong glue made from two resins and individually neither can make the glue sticky. While putting those two resins together, a strong and powerful adhesive is obtained because a chemical bond is produced as the inset of Fig. 1a shows. Inspired by the epoxy, we experimentally report a recipe for fully controlling SH photons via the proposed integrated nonlinear optical interfaces. As a proof-of-concept validation, Fig. 1a illustrates a schematic of bonding $WS_2$ monolayer with linear plasmonic Au nanosieve, which behaves as a hybrid nonlinear optical interface for multifunctional SHG manipulation. In this hybrid, SHG is in principle forbidden in Au nanosieves made of centrosymmetric rectangular nanoholes, but the whole integrated optical interface offers highly efficient SHG emission, whose properties may be precisely controlled by hundred-nanometer pixel-level resolution. SHG, originated from the monolayer $WS_2$, can be enhanced by strong local-field within each Au nanoholes, and simultaneously the geometric phase determined by individual Au nanoholes could be locally engineered and imparted to the $WS_2$ monolayer on top of the Au nanosieve. The experiment result indicates that a large effective SHG susceptibility of ~25 nm/V is achieved, which is 3-order of magnitude larger than that of conventional plasmonic metasurfaces. Multifunctional nonlinear optical devices have been demonstrated in a series of free-space transmissive nonlinear optical interfaces for orbit angular momentum (OAM) generation, beam steering, versatile polarization control and holograms and as shown in Fig. 1a. This strategy could readily lead to the realization of a vast plethora of designer functions. We firmly believe that the work reported here may provide a distinct platform, i.e., TMDC-metasurface nonlinear optical interfaces, to realize robust nonlinear applications (e. g. high-order harmonic generation, HHG) with high efficiency, ultra-compactness and flexible fabrication.

Fig. 1b shows the atomic force microscope (AFM) image for the WS$_2$ flake and the height profile is measured to be ~0.7 nm, indicating that a monolayer WS$_2$ is used. An optical image of monolayer WS$_2$ on the Au nanosieve is presented in Fig. 1c, showing the Au metasurface is fully covered by a triangular monolayer WS$_2$ with an area of $100~\mu m^2$ (See Fig. S1 in SI). Fig. 1d presents a scanning electron microscopy (SEM) image of the same sample indicated by the square area in Fig. 1c, where the high-quality WS$_2$ monolayer can be clearly observed. Since the Au nanosieve acts as a phase plate for the fundamental laser, the SPR peak of the rectangular Au nanohole should be matched with the wavelength of the fundamental laser at 810 nm for a large modulation efficiency. In our experiment, the Au nanohole has the length of 185 nm, the width of 80 nm, and the thickness of 60 nm. The effective second-order susceptibility of the Au-WS$_2$ interface is estimated to be ~25 nm/V, which is three-order of magnitude larger compared with the one of typical plasmonic metasurfaces (~ pm/V) (See Fig. S2 in SI for details). Fig. 1e presents the two-photon induced photoluminance (TPL) spectrum from a monolayer WS$_2$ under an excitation at 810 nm with a peak wavelength at ~620 nm. Fig. 1f shows the Raman spectrum excited at 532 nm from a monolayer WS$_2$. Two typical signatures located at 352.3 and 417.3 cm$^{-1}$ can be observed, which are attributed to the $E_{2g}^1$ and $A_g^1$ phonon modes from the monolayer WS$_2$, respectively. In general, the characterization results are in good agreement with the previous reports, indicating the high quality of the monolayers WS$_2$ and its intrinsic properties are not changed in the hybrid Au-WS$_2$ interface, which are essentially responsible for the outstanding results in the work.

The underlying physics of SHG phase control in the integrated Au-WS$_2$ interface are illustrated in Fig. 2a. For a single Au nanohole, the rotation angle, $\theta$, is defined as the angle between the long-axis of the rectangular nanohole and *x*-axis. When a linearly polarized (LP) fundamental beam (polarized along *y*-axis), $\sigma^{(0,\omega)}$, is normally incident on a rectangular Au nanohole, the transmitted beam can be decomposed into three components (see section 1 in SI for detailed discussions): left circularly polarized (LCP, $\sigma_-$) and right circularly polarized (RCP, $\sigma_+$) beams with geometric phase discontinuity of $\pm 2\theta$, and a LP beam without geometric phase, which are written as, $e^{(\mp i2\theta)}\sigma^{(\pm,\omega)}$ and $\sigma^{(0,\omega)}$, accordingly. As these fundamental beams further pump the monolayer WS$_2$, the polarization distribution of the SH signals from the monolayer WS$_2$ are illustrated (suppose crystalline *a*-axis of monolayer WS$_2$ is along *x*-axis). Due to the D$_{3h}$ symmetry of monolayer WS$_2$, the emitted SH signal maintains LP while the polarization directions are rotated by $2\theta$. Furthermore, the emitted SH in the same nanohole has five components with different phase jump and polarization. The 0$^{th}$ order SHG beam is a LP beam without deflection (along the *z*-axis). The $\pm$1$^{st}$ order and $\pm$2$^{nd}$ order deflected beams are CP beams (LCP and RCP) with a phase delay of $\mp 2\theta$ and $\mp 4\theta$, respectively. Due to the conservation of angular momentum, these deflected SHG beams have an orthogonal polarization state to the excitation fundamental beams. Therefore, the $\pm$1$^{st}$ order and $\pm$2$^{nd}$ order deflected beams can be written as, $e^{(\pm 2i\theta)}\sigma^{(\pm,2\omega)}$ and $e^{(\pm 4i\theta)}\sigma^{(\pm,2\omega)}$, respectively. It indicates that the SHG phase can be precisely controlled by orientating Au nanoholes, which are also entangled with the spin angular momentum (SAM). Note that the intensity of $\pm$1$^{st}$ order deflected beams are much stronger than that of the $\pm$2$^{nd}$ order beams, so we mainly focus on the manipulation of $\pm$1$^{st}$ order SHG beams in the following. Lastly and importantly, note the mechanisms as well as the results demonstrated here are fundamentally different from the conventional nonlinear metasurfaces made of special symmetric nanostructures (See section 2 in SI for further discussions in details), which highlights our novel perspectives of hybridization of linear plasmonic metasurface with monolayer semiconductors with giant nonlinearity.

For demonstrating the mechanism of SHG manipulation, we design a linear plasmonic nanosieves with a spiral phase and the gradient phase as shown in the SEM image in Fig. 2b. (See Fig. S3 in SI for details). It shows a patterned rectangular Au nanoholes array with varying orientations, and a hexagonal arrangement with a period of 400 nm is employed to increase the density of unit cells. Specifically, Fig. 2c shows the polarization distribution of the fundamental beam passed through the Au nanohole array (indicated by the rectangle area in Fig. 2b), which are majorly polarized along the short-axis of each nanohole (indicated by the arrows). The deeper color indicates a stronger local-fields in the Au nanohole. The inset of Fig. 2d presents the phase distribution of the 1$^{st}$-order of RCP component of the SH emission from each cell under a LP pumping with a polarization along $y$-axis, which can be described as,

$$x \operatorname{Sin}(\alpha)/\lambda + n\gamma/2\pi, \tag{1}$$

where $x$ is the $x$-coordinates of nanoholes, $\lambda$ is the wavelength of SHG, $\alpha$ is the deflection angle, $n$ is the topological charge, and $\gamma$ is the azimuth angle in circular coordinate system. Here, the deflection angle is $\alpha = 10°$ and the geometric topological charge is $n = 1$. Note that the area indicated by square is in accordance with the SEM image in Fig. 2b. In Fig. 2d, it reports the measured spatial intensity profiles of emitted SH signals under a LP fundamental beam ($y$-axis polarized), and five SHG beams are clearly seen as predicted by our theory. Specifically, the 0$^{th}$ order beam with a circular spot is the LP beam without deflection. The $\pm 1^{st}$ order SH beams are propagated along both sides of the $z$-axis with the same defection angle, indicating a doughnut-shaped spot due to the properties of OAM. Both beams have an opposite CP with the same value of deflection angle, $\alpha = 10°$, which is consistent with the design. Importantly, the topological charge is determined to be $\pm 1$ by interference pattern between 0$^{th}$ order and $\pm 1^{st}$ order beams (the details for the interference patterns is described in Figure S4 in SI). Note that both $\pm 2^{nd}$ order SHG beams present a thinner doughnut-shaped pattern with a much weaker intensity than the $\pm 1^{th}$ order SHG beams, whose topological charge is $\pm 2$ in theory.

The deflection angle is measured to be around 20°, which is twice the one of the $\pm 1^{st}$ order beams under the small deflection angle approximation.

Interestingly, the phase plate of SH emission of the same sample is appeared quite different under excitation of a CP fundamental beam, as shown in the inset of Fig. 2e. Since the phase delay is doubled under CP excitation in comparison with LP excitation (see Fig. S3 in SI for details). Fig. 2e shows only two SHG beams: the SHG beam along the $z$-axis is LP without deflection, similar with the $0^{th}$ order in Fig. 2d. The $-1^{st}$ order deflected SHG with a deflection angle of 20° whose intensity distribution is doughnut-shaped with topological charge of -2, consistent with our theoretical expectations. The spot shape, polarization, deflected angle and topological charge of the $-1^{st}$ order deflected SHG beam are the same with the $-2^{nd}$ order SHG beam under LP excitation shown in Fig. 2d. Such difference can be attributed to the dependence of SHG on the amplitude of the fundamental beam, as further discussed in SI. Our demonstration of dynamic nonlinear OAM generation under excitation with different polarizations could serve as the nonlinear version of the recently demonstrated metasurface-enabled quantum entanglement of SAM and OAM[37]. Meanwhile, the change of spatial deflection angle of $-1^{st}$ order in different pumping polarization, i.e. resulting in different spots in the image plane, can also be potentially further explored to realize the nonlinear quantum metasurface for photon state reconstruction[38]. Thus, our demonstration of spin-entangled and spatially tunable nonlinear OAM here could be very powerful in free-space quantum imaging, generation of entangled photon states and other applications. Overall, these results provide clear evidence of the hundred-nanometer-precision phase control of SHG from the Au-WS$_2$ interface.

In what follows, we further elaborate how this strategy could be implemented to realize multi-functional nonlinear optical devices. In the inset of Fig. 3a, it shows the phase distribution of the SH signals of a new sample, analog to a typical photonic spin-Hall metasurface with a geometric phase gradient ($\nabla \varphi$) along $x$-axis.[39, 40] Fig. 3a presents the measured propagation of emitted SHG along the z-axis from the sample. The emitted SH signals are split into three beams. The inset images captured at different positions along the z-axis present a clear illustration of the SHG spots, indicating a low divergence angle of the SHG beams. Specifically, the $0^{th}$ order SHG beam propagates along the z-axis without any deflection due to the fundamental laser without carrying geometric phase, while the rest travel along two sides of the z-axis with the same defection angle ($\pm 1^{st}$ order). The deflection angle $\alpha$ is measured to be 10°, in a good agreement with the theoretical designs (10°). Furthermore, the polarization states of the SH beams are analyzed. The Stokes parameters $S_3$ ($S_3 = \frac{I_{RCP} - I_{LCP}}{I_{RCP} + I_{LCP}}$) of -1, 0, 1-order SH beams are determined to be -0.90, -0.32, and 0.95, respectively. This indicates that the $\pm 1^{st}$ order deflected beams have opposite CP with almost pure polarization, while the $0^{th}$ order SH beam is approximately LP. In Fig. 3b-3d, we show the extracted RCP component from the experimental results from different samples with deflection angles of 5°, 10° and 15°, respectively. Correspondingly, the deflected RCP beams observed at $\alpha$=5°, 10° and 15°, while the RCP component of the $0^{th}$ order beam propagates without deflection and the LCP component is blocked by a quarter-wave plate coupled with a glan-laser polarizer. Interestingly, a weak $2^{nd}$ order SH beam can also be observed with a deflection angle of ~10° (Fig. 3b), which is approximately twice of the design angle of 5°. However, it cannot be observed in Fig. 3c and 3d, which may be ascribed to a low diffraction efficiency at large defection angle and the limited detection efficiency in experiment (see Fig. S5 and S6 in SI).

Following the beam steering results, two deflected SHG beams propagate along different sides of the z-axis with opposite spin angular momentums (SAM). By reversing the order of the supercells arranged on the Au nanosieve, the SAM of the deflected SHG beams can be reversed, while keeping the same defection angle. Then, this integrated interface can generate a LP beam by a superposition of two SHG beams with opposite SAMs, whose polarization can be controlled arbitrarily by adjusting the phase difference of the two SHG beams. Fig. 3e illustrates the schematic of versatile polarization control of the deflected SHG beam under excitation with a LP fundamental laser (y-axis polarized).

Specifically, by reversing the direction of phase gradient of the supercells arranged on the odd lines of the Au metasurface, half are converted to deflect an opposite SAM while keeping the same defection angle ($\alpha=10°$), and thus we can generate a LP beam. Note that both deflected LP beams have the same polarization. The phase difference of $\sigma^{+,2\omega}$ and $\sigma^{-,2\omega}$ can be controlled by adjusting the optical path difference, $d\mathrm{Sin}(\alpha)$, where $d$ is the offset of the odd and even lines of the metasurfaces with an inversed order of the supercells. Figure 3f and 3g present the polarmetric plots of the measured SHG intensity as a function of polarization angle, $\beta$, at phase differences, $d\mathrm{Sin}(\alpha) = 0\lambda, 0.5\lambda$, respectively. $\beta = 0°$ implies that the SHG beam is linearly polarized along y-axis. All measured SHG beams are approximately pure linearly polarized. The polarization directions for each beam are indicated by the long axis of the patterns at $10°, 100°$, respectively. This is in good agreement with the design rule $\beta = \frac{d\,\mathrm{Sin}(\alpha)}{\lambda}\pi$, where $\lambda$ is the SHG wavelength. We find the same deviation angle of $10°$ for each SHG beam, which is attributed to the orientation of the WS$_2$ monolayer[28]. This is quite different from the polarization control in the linear case[41]. Due to the D$_{3h}$ symmetry of monolayer WS$_2$, the polarization direction of the emitted SH signal is rotated by $3\varphi$ as the monolayer WS$_2$ is rotated by $\varphi$. The rotation angle of the monolayer WS$_2$ is measured to be $\varphi = 3°$, according to the optical microscopy image in Fig. S1b[28], leading to a deviation angle of $10°$. Such versatile nonlinear polarization generation could find applications in commercial polarization modulators[41].

Such nonlinear integrated interface also allows holographic imaging, as schematically illustrated in Figure 4a, where the circle with a diameter of 30 μm indicates the Au nanohole array. Under a LCP fundamental beam, RCP component of emitted SHG beam from monolayer WS$_2$ can be imaged in the Fresnel region. The image plane is designed to be 200 μm away from the sample, and each letter in the pattern is 25 μm wide. The designed objective images are the short form for "Huazhong University of Science and Technology" and "National University of Singapore" (Fig. 4b). The phase plane can be generated by iterative method (further details are available in the Fig. S7 in SI). The theoretical images calculated by Fresnel-Kirchhoff's diffraction formula and the measured experimental images are shown in Fig. 4c and 4d, respectively. The intensity distributions of the measured images match with the theory very well.

In conclusion, inspired by the epoxy, we proposed and experimentally demonstrated a new platform for full control of the nonlinear wavefront of SHG based on Au-TMDC nonlinear optical interface. It fuses Au nanosieve with a monolayer $WS_2$, and the former acts as an unprecedentedly strong and precise "spatial light modulator" down to the pixel size of hundred-nanometer for the latter. Our results prove that multi-functional SHG control, such as SH beam steering, versatile polarization control, dynamic OAM generation and holography are enabled, together with large effective second-order nonlinear susceptibility of ~25 nm/V. It provides a direct evidence of both efficiently generating and manipulating SHG signals with this strategy. From the application viewpoint, this control enables miniaturized nonlinear optical devices with good performance in the scale of tens of nanometers in thickness. It therefore opens up a wide range of opportunities to realize ultrathin, ultra-compact nonlinear optical devices, such as beam splitting, versatile polarization control, free-space quantum communications and others.

## Methods

**Sample fabrication.** (1) The monolayers $WS_2$ were fabricated on a sapphire substrate (6-Carbon Tech.). The solution of polymethyl methacrylate (PMMA, 2 wt%, Aldrich) was drop-coated on the $WS_2$-sapphire substrate. It was placed at room temperature for 1.5 h, and then baked at 120 °C for 0.5 h. The $WS_2$-sapphire substrate was immersed in NaOH aqueous solution (3 mol/L) at room temperature for more than 4 h to etch the sapphire surface, and a PMMA-$WS_2$ films can be naturally lifted off from the sapphire substrate. The PMMA-$WS_2$ films were fished out with a glass slide and immersed in deionized water for 3 times in order to remove the residual NaOH solution; (2) A 60-nm thick gold film with a 5-nm thick Cr adhesion layer were deposited on a quartz substrate by using e-beam evaporator. Then, the Au nanoholes array with different arrangement were milled by the focused ion beam milling method. (FIB, FEI Versa 3D); (3) With the help of motorized stages controlled needle, the PMMA-WS$_2$ films were fished out by the fabricated metasurfaces, and then it must be aligned quickly with the Au metasurface under the microscope before the water was dry. Finally, it was baked at 120 °C for 1 h to improve the combination of WS$_2$ monolayers and the Au nanosieve and immersed in acetone for 3 times to remove the PMMA film.

**SHG characterization.** A mode-locked Ti-sapphire femtosecond laser centered at 810 nm (Vitara Coherent, 8 fs and 80 MHz) was used for the fundamental beam source. The polarization was adjusted by a quarter-wave plate (WPQ05M-808, Thorlabs). A home-built optical system was used to measure emitted SH signal under an excitation of a plane-wave laser. The fundamental beam was focused by an 8-cm lens or an objective (Olympus, 10× and 0.25 NA) to generate different sizes of focal spots. The emitted signal was collected by an objective lens (Olympus, 40× and 0.65 NA), filtered spectrally, and imported to a CMOS camera (Prime 95B, PHITIMETRICS) or to a spectrometer (Acton 2500i with Pixis CCD camera, Princeton Instruments) through a fiber. The emitted SH signal was extracted by a Glan-laser polarizer with a quarter-wave plate at the SH wavelength (GL10-A and WPQ05M-405, Thorlabs). To measure the spatial intensity of the SH signals, we captured the SHG images at the different planes from $0\ \mu m$ to $200\ \mu m$ along $z$-axis with a step-size of $0.5\ \mu m$.

## Acknowledgment

This work was supported by National Natural Science Foundation of China (nos. 91850113 and 11774115), the 973 Programs under grants 2014CB921301, the Doctoral fund of Ministry of Education of China under Grant No. 20130142110078. Special thanks to the Center of Nano-Science and Technology of Wuhan University for their support of sample fabrication.

## Author contributions

K. W., P.X.L. and C.W.Q. conceived the project. K.W., P.X.L., A.A., and C.W.Q. supervised the project. K.W., G.W.H., A.A., and C.W.Q. designed the experiments. X.M.H., W.C.Z., K.W. and W.W.L. performed the experiments. G.W.H., K.W., B.W., P.X.L. A.A. and C.W.Q. analyzed data. All authors discussed the results. G.W.H. and K. W. drafted the paper with the inputs from all authors.

## Additional information

Supporting information is available.

## Competing financial interests

The authors declare no competing financial interests.

## Figure Captions

**Fig. 1 | Schematic illustration and material characterizations of the epoxy-inspired Au-TMDC nonlinear optical interface. a.** Sketch of the epoxy-inspired Au-TMDC nonlinear optical interface. It offers highly efficient SHG emission from monolayer $WS_2$ whose properties are precisely controlled by pixel-level resolution, opening the door for many nonlinear beam-shaping applications, some of which are presented in the insets; **b.** AFM image for a monolayer $WS_2$ with a height profile of ~0.7 nm. **c.** Optical image of the Au-$WS_2$ interface. The Au metasurface is fully covered by a triangular monolayer $WS_2$. **c** and **d**. TPL and Raman spectrum from a monolayer $WS_2$, respectively.

**Fig. 2 | Illustration of the principle of SHG phase control in the integrated Au-$WS_2$ interface and OAM generation under LP and CP excitation. a.** Illustration of the principle of SHG phase control in the Au-WS2 interface. **b.** SEM image for a linear plasmonic nanosieves with a spiral phase and the gradient phase. **c.** Polarization distributions (indicated by the arrows) of the fundamental beam passed through the Au nanohole array in the area indicated by the square in **b**. The deeper color indicates

stronger local-fields in the Au nanoholes. d. Measured spatial intensity (Log) profile of emitted SHG beams. The inset indicates the SHG phase profile of the sample for OAM generation under the LP fundamental beam (polarized along *y*-axis). e. Measured spatial intensity profile of emitted SHG beams. The inset indicates the SHG phase profile of the same sample under the CP fundamental beam; Note that the area indicated by square is in accordance with the SEM image shown in **b**.

**Fig. 3 | Demonstration of SHG beam steering and versatile polarization generation. a.** Experimentally measured propagation of emitted SH signal along *z*-axis under a LP excitation (polarized along *y*-axis). The inset shows the phase distribution of the SH signals of the new sample, analog to a typical photonic spin-Hall metasurface with a geometric phase gradient ($\nabla\varphi$) along *x*-axis. The inset images captured the different *z*-axis present clearly illustration of the emitted SH spots. **b-d.** measured RCP component from samples with different deflected angle of 5°, 10° and 15°, respectively. **e.** Schematic of versatile control of the polarization of the deflected LP beams. **f** and **g.** Polarimetric plots of SHG intensity from two samples with phase differences of $d\mathrm{Sin}(\alpha)=0\lambda$, $0.5\lambda$, respectively. It indicates that the emitted SH signals have almost pure linear polarization.

**Fig. 4 | Demonstration of SHG holography. a.** Schematic illustration of holographic imaging of SHG beam under a LCP fundamental beam in the Fresnel region; **b-d.** The objective, theoretical and experimental images of holographic imaging.

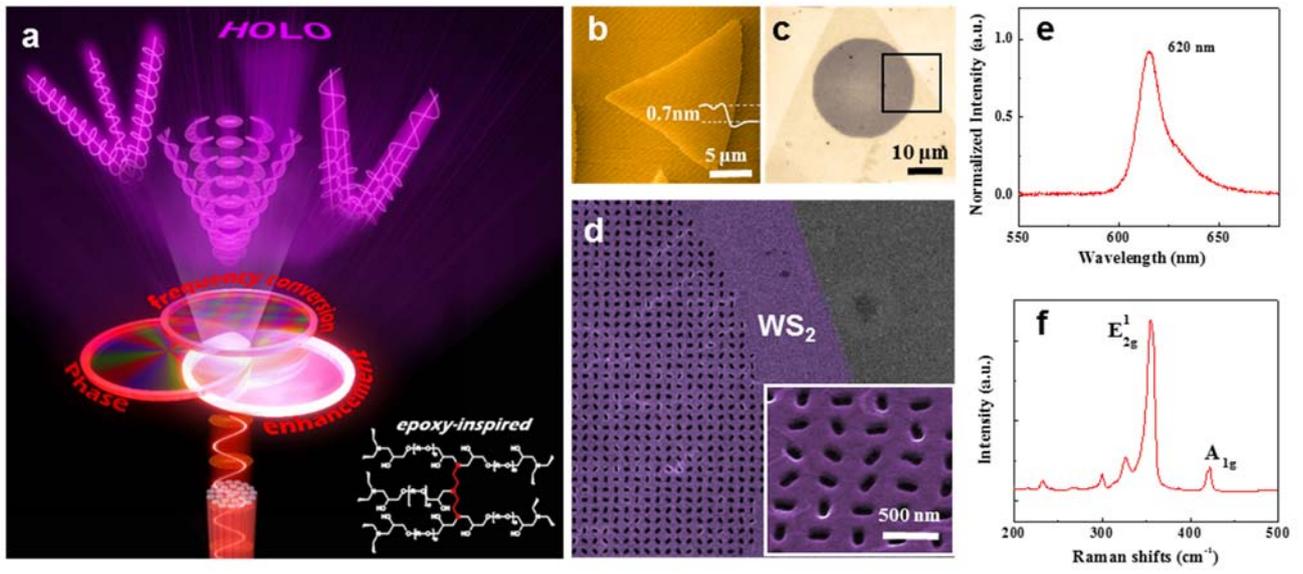

**Figure 1**

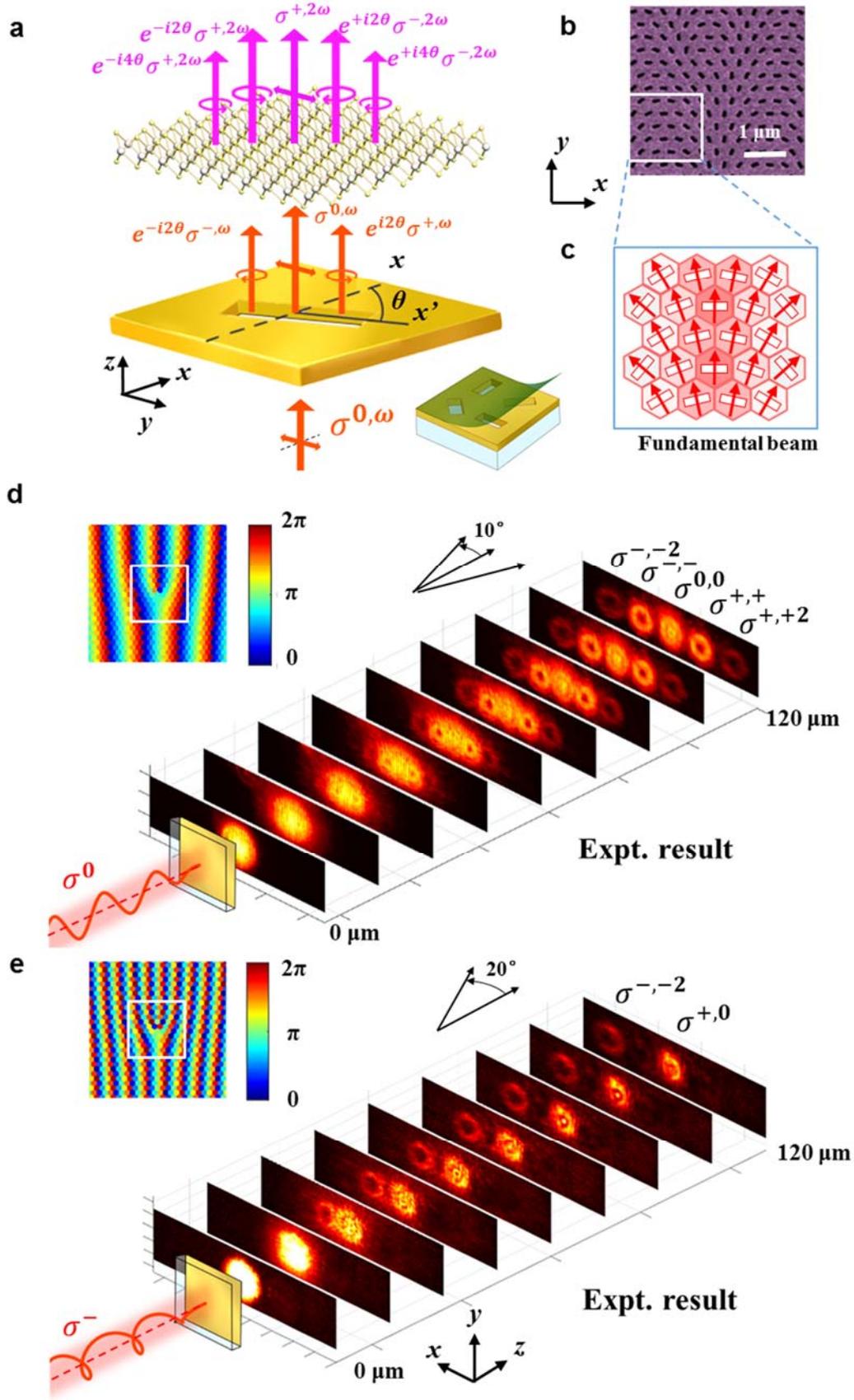

**Figure 2**

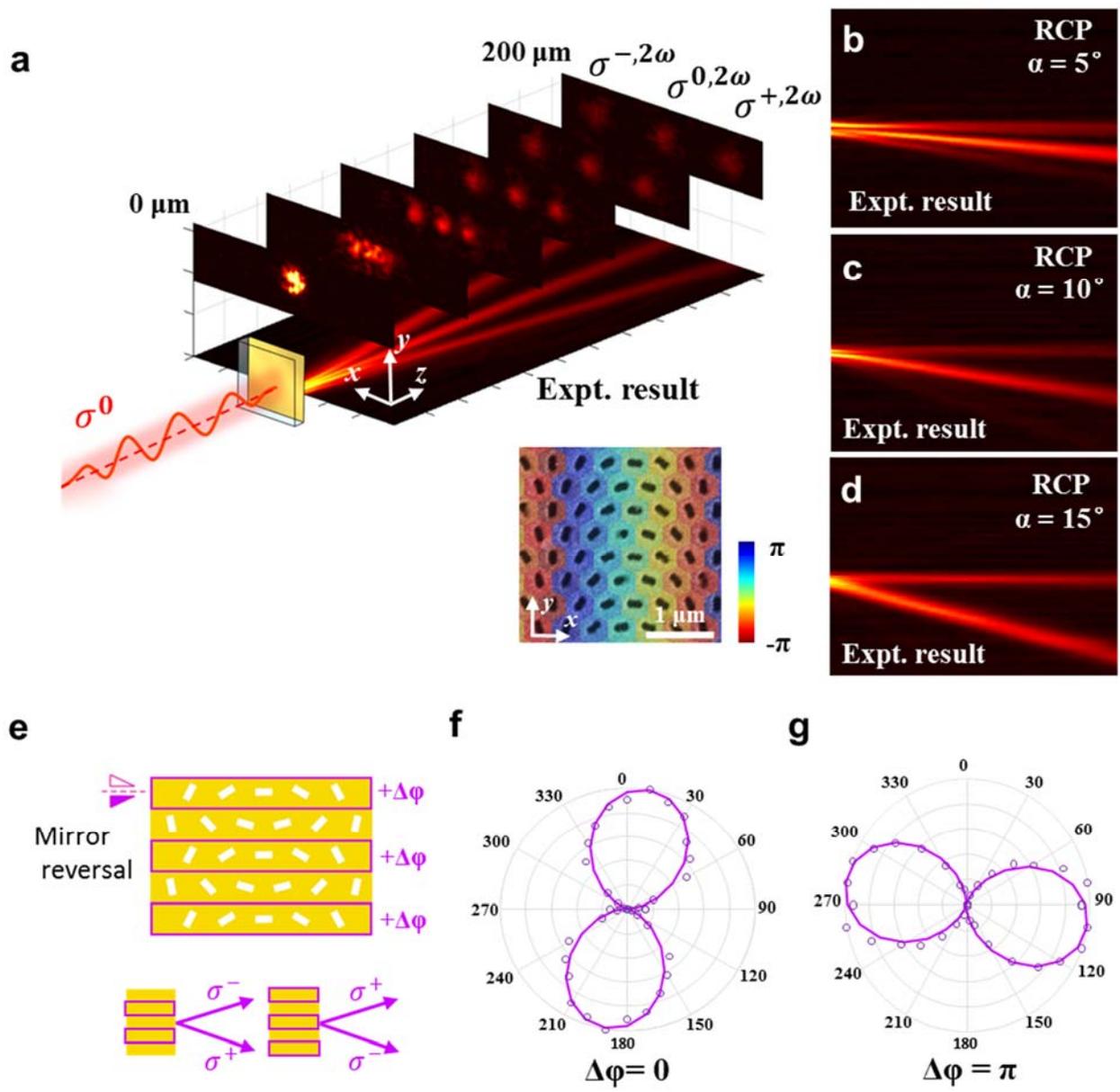

**Figure 3**

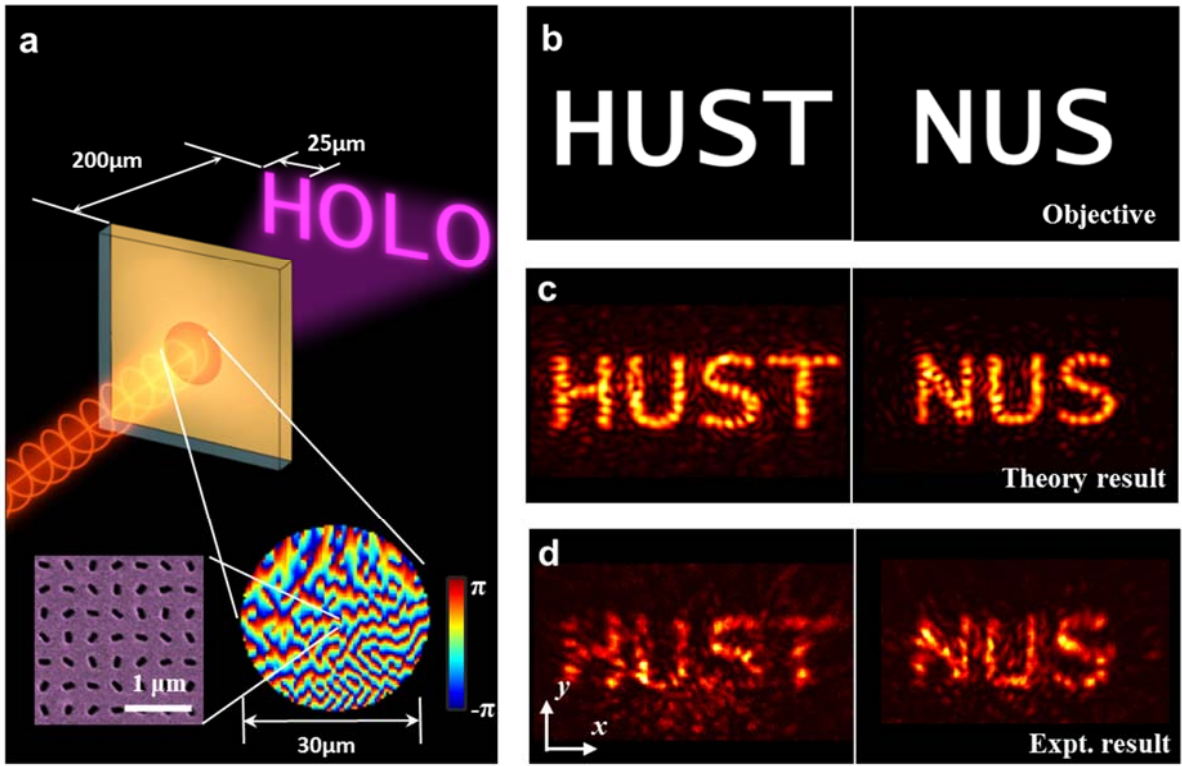

**Figure 4**